\documentclass[runningheads]{svmult}

\usepackage{makeidx}

\usepackage{graphicx}

\usepackage{subeqnar}

\usepackage{multicol}

\usepackage{cropmark}

\usepackage{physprbb}

\def\leti{Lense--Thirring}
\def\zone{the error due to the even zonal harmonics of the geopotential\ }
\def\bm#1{{\mbox{\boldmath$#1$\unboldmath}}}
\def\rfr#1{(\ref{#1})}

\def\eqi{\begin{equation}}
\def\eqf{\end{equation}}
\def\eqia{\begin{eqnarray}}
\def\eqfa{\end{eqnarray}}

\def\lb#1{\label{#1}}

\begin{document}

\title*{New perspectives in testing the general relativistic Lense--Thirring effect}

\toctitle{New perspectives in testing  \protect\newline the
general relativistic Lense--Thirring effect}

\titlerunning{New perspectives in testing the general relativistic Lense--Thirring effect}

\author{Lorenzo Iorio\inst{1} }

\authorrunning{Lorenzo Iorio}

\institute{Dipartimento di Fisica dell'Universit${\rm \grave{a}}$
di Bari, via Amendola 173, 70126, Bari, Italy}

\maketitle

\begin{abstract}
Testing the effects predicted by the General Theory of Relativity,
in its linearized weak field and slow motion approximation, in the
Solar System is difficult because they are very small. Among them
the post-Newtonian gravitomagnetic Lense-Thirring effect, or
dragging of the inertial frames, on the orbital motion of a test
particle is very interesting and, up to now, there is not yet an
undisputable experimental direct test of it. Here we illustrate
how it could be possible to measure it with an accuracy of the
order of 1$\%$, together with other tests of Special Relativity
and post-Newtonian gravity, with a joint space based OPTIS/LARES
mission in the gravitational field of Earth. Up to now, the data
analysis of the orbits of the existing geodetic LAGEOS and LAGEOS
II satellites has yielded a test of the Lense-Thirring effect with
a claimed accuracy of 20$\%$-30$\%$.

\end{abstract}

\section{Introduction}
The linearized weak--field and slow--motion approximation of the
General Theory of Relativity (GTR) \cite{ciuwhe95} is
characterized by the condition $g_{\mu\nu}\sim
\eta_{\mu\nu}+h_{\mu\nu}$ where $g_{\mu\nu}$ is the curved
spacetime metric tensor, $\eta_{\mu\nu}$ is the Minkowski metric
tensor of the flat spacetime of Special Relativity and the
$h_{\mu\nu}$ are small corrections such that $|h_{\mu\nu}| \ll 1$.
Until now, many of its predictions, for the motion of light rays
and test masses have been tested, in the Solar System, with a
variety of techniques to an accuracy level of the order of $0.1\%$
\cite{wil93, wil01}. It is not so for the
gravitomagnetic\footnote{In the weak field and slow motion
approximation of GTR the equations of motion of a test particle
freely falling in the gravitational field of a central spinning
body are formally analogous to those governing the motion of an
electrically charged particle in an electromagnetic field under
the action of the velocity--dependent Lorentz force. In the
gravitational case the role of the magnetic field is played by the
so called gravitomagnetic field which is generated by the
off--diagonal terms of the metric $g_{0i}$ and whose source is the
proper angular momentum \bm J of the central body.}
Lense--Thirring effect due to its extreme smallness. It can be
thought of as a consequence of a gravitational spin--spin
coupling.

If we consider the motion of a spinning particle in the
gravitational field of a central body of mass $M$ and proper
angular momentum \bm J,  it turns out that the spin \bm s of the
orbiting particle undergoes a tiny precessional motion
\cite{sch60}. The most famous experiment devoted to the
measurement, among other things, of such gravitomagnetic effect in
the gravitational field of Earth is the Stanford University GP--B
mission \cite{eveetal01} which should fly at the end of 2003, in
spite of recent problems \cite{law03a,law03b}.

If we consider the whole orbit of a test particle in its geodesic
motion around $M$ as a sort of giant gyroscope, its orbital
angular momentum \bm \ell\ undergoes the Lense--Thirring
precession, so that the longitude of the ascending node $\Omega$
and the argument of pericentre $\omega$ of the orbit\footnote{The
longitude of the ascending node $\Omega$ is an angle in the
reference $\{x,y\}$ plane, which usually coincides with the
equatorial plane of the central body, counted from the reference
$x$ axis to the line of the nodes. The line of the nodes is given
by the intersection between the orbital plane and the reference
plane. The argument of the pericentre $\omega$ is an angle in the
orbital plane counted from the line of the nodes to the pericentre
of the orbit which is the point of closest approach of the
orbiting particle to the central body. In the original paper by
Lense and Thirring the longitude of the pericentre
$\varpi=\Omega+\omega$ is used instead of $\omega$. }  of the test
particle \cite{ste60} are affected by tiny secular precessions
$\dot\Omega_{\rm LT}$, $\dot\omega_{\rm LT}$ \cite{leti18,
ciuwhe95, ior01}.
%----------------------------------------------------------------------
\subsection{The LAGEOS-LAGEOS II Lense-Thirring experiment}
Up to now, the only attempts to detect the \leti\ effect on the
orbit of test particles in the gravitational field of Earth are
due to Ciufolini and coworkers \cite{ciuf00} who analysed the
laser data of the existing geodetic passive SLR (Satellite Laser
Ranging) satellites LAGEOS and LAGEOS II over time spans of some
years. The observable is a suitable combination of the orbital
residuals of the nodes of LAGEOS and LAGEOS II and the perigee of
LAGEOS II according to an idea exposed in \cite{ciuf96}. The
relativistic signal is a linear trend with a slope of almost 60.2
milliarcseconds per year (mas yr$^{-1}$ in the following). The
standard, statistical error is evaluated as 2$\%$. The claimed
total accuracy, including various sources of systematical errors,
is of the order of $20\%-30\%$.

The main sources of systematical errors in this experiment are
\begin{itemize}
\item
the unavoidable aliasing effect due to the mismodelling in the
classical secular precessions induced on $\Omega$ and $\omega$ by
the even zonal coefficients $J_{l}$ of the multipolar
expansion\footnote{It accounts for the oblateness of Earth
generated by its diurnal rotation.} of the terrestrial
gravitational field \cite{kau66}
\item
the non--gravitational perturbations affecting especially the
perigee of LAGEOS II \cite{luc01, luc02}. Their impact on the
proposed measurement is difficult to be reliably assessed
\end{itemize}
It turns out that the mismodelled classical precessions due to the
first two even zonal harmonics of the geopotential $J_2$ and $J_4$
are the most insidious source of error for the Lense--Thirring
measurement with LAGEOS and LAGEOS II. The combination of
\cite{ciuf96} is insensitive just to $J_2$ and $J_4$. According to
the full covariance matrix of the EGM96 gravity model
\cite{lemetal98}, the error due to the remaining uncancelled even
zonal harmonics amounts to almost 13$\%$ \cite{iorcelmec03}.
However, if the correlations among the even zonal harmonic
coefficients are neglected and the variance matrix is
used\footnote{Such approach is considered more realistic by some
authors \cite{riesetal98} because nothing assures that the
correlations among the even zonal harmonics of the covariance
matrix of the EGM96 model, which has been obtained during a
multidecadal time span, would be the same during an arbitrary past
or future time span of a few years as that used in the
LAGEOS--LAGEOS II
\leti\ experiment or in the proposed LAGEOS--LARES mission.},
\zone amounts to 46.6$\%$ \cite{iorcelmec03}. With this estimate
the total error of the LAGEOS--LAGEOS II
\leti\ experiment would be of the order of 50$\%$.
%---------------------------------------------------------------------------
\subsection{The LARES project}
The originally proposed LAGEOS III--LARES mission \cite{ciuf86}
consists of the launch of a LAGEOS--type satellite--the
LARES--with the same orbit of LAGEOS except for the
inclination\footnote{The inclination $i$ is the angle between the
orbital plane and the reference plane. It is counted from the
reference plane so that equatorial orbits have $i=0$ deg. The semi
major axis $a$ and the eccentricity $e$ fix the size and the
shape, respectively, of the orbit of the test particle. For closed
orbits $0<e<1$. Circular orbits have $e=0$.} $i$ of its orbit,
which should be supplementary to that of LAGEOS, and the
eccentricity $e$, which should be one order of magnitude larger in
order to perform other tests of post--Newtonian gravity
\cite{iorciufpav02, ioryuk02}. In Table 1 the orbital parameters
of the existing and proposed LAGEOS--type satellites are quoted.
%...........................................................................

\begin{table}
\caption{Orbital parameters of LAGEOS, LAGEOS II and LARES and
their Lense--Thirring precessions.}
\begin{center}
\renewcommand{\arraystretch}{1.4}
\setlength\tabcolsep{5pt}
\begin{tabular}{lllll}
\hline
\noalign{\smallskip}
Orbital parameter & LAGEOS & LAGEOS II & LARES\\
\noalign{\smallskip}
\hline
\noalign{\smallskip}
$a$ semi major axis (km) & 12270 & 12163 & 12270\\
$e$ eccentricity & 0.0045 & 0.014 & 0.04\\
$i$ inclination (deg) & 110 & 52.65 & 70\\
$\dot\Omega_{\rm LT}$ (mas yr$^{-1}$) & 31 & 31.5 & 31\\
$\dot\omega_{\rm LT}$ (mas yr$^{-1}$)& 31.6 & -57 & -31.6\\
\hline
\end{tabular}
\end{center}
\label{table1}
\end{table}

%.....................................................................................
The choice of the particular value of the inclination for LARES is
motivated by the fact that in this way, by using as observable the
sum of the nodes of LAGEOS and LARES, it should be possible to
cancel out to a very high level all the contributions of the even
zonal harmonics of the geopotential, which depends on $\cos i$,
and add up the Lense--Thirring precessions which, instead, are
independent of $i$. The use of the nodes would allow to reduce
greatly the impact of the non--gravitational perturbations to
which such Keplerian orbital elements are rather insensitive
\cite{luc01, luc02}.

Of course, it would not be possible to obtain practically two
orbital planes exactly 180 deg apart due to the unavoidable
orbital injection errors which can be considered of the order of
0.5--1 deg. In Figure 1, page 4314 of \cite{iorlucciuf02} and
Figure 1, page 1267 of \cite{ior03} the impact of such source of
error on the originally proposed LAGEOS--LARES mission has been
shown. It could amount up to 4$\%$ for an injection error of 1 deg
in the inclination of LARES.

In \cite{iorlucciuf02} an alternative observable based on the
combination of the residuals of the nodes of LAGEOS, LAGEOS II and
LARES and the perigee of LAGEOS II and LARES has been proposed. It
would allow to cancel out the first four even zonal harmonics so
that the error due to the remaining even zonal harmonics of the
geopotential would be rather insensitive both to the orbital
injection errors in the LARES inclination and to the correlations
among the even zonal harmonic coefficients. It would amount to
$0.02\%$--$0.1\%$ only \cite{iorlucciuf02, ior03}.

In regard to the present status of the LARES project,
unfortunately, up to now, although its very low cost with respect
to other much more complex and expensive space--based missions, it
has not yet been approved by any national space agency or
scientific institution.
%---------------------------------
%\begin{figure}[b]

%\begin{center}

%\includegraphics[width=.3\textwidth]{kepell.eps}

%\end{center}

%\caption[]{Keplerian orbital elements of a test particle $m$
%moving in the gravitational field of a central rotating body of
%mass $M$ and proper angular momentum $J$.  }

%\label{kepelems}

%\end{figure}
%--------------------------------------------------------
\section{The proposal of a joint OPTIS/LARES mission}
\subsection{The originally proposed OPTIS mission}
OPTIS \cite{lametal01} is a recently proposed satellite--based
mission\footnote{See also on the WEB
http://www.exphy.uni-duesseldorf.de/OPTIS/optis.html.} which would
allow for much improved tests of
%..................................................
\begin{itemize}
\item
the isotropy of the velocity of light
\item
the independence of the velocity of light from the velocity of the
laboratory
\item
the universality of the gravitational redshift.
\end{itemize}
%....................................................
This mission is based on the use of a spinning drag--free
satellite in an eccentric, high--altitude orbit which should allow
to perform a three orders of magnitude improved Michelson--Morley
test and a two order of magnitude improved Kennedy--Thorndike
test. Moreover, it should also be possible to improve by two
orders of magnitude the tests of the universality of the
gravitational redshift by comparison of an atomic clock with an
optical clock. The proposed experiments are based on ultrastable
optical cavities, lasers, an atomic clock and a frequency comb
generator. Since it is not particularly important for the present
version of the mission, the final orbital configuration of OPTIS
has not yet been fixed; in \cite{lametal01} a perigee
height\footnote{With respect to Earth's surface.} of 10000 km and
apogee height of 36000 km are provisionally proposed assuming a
launch with a Ariane 5 rocket.

The requirements posed by the drag--free technology to be used,
based on the field emission electrical propulsion (FEEP) concept,
yield orbital altitudes not less than 1000 km. On the other hand,
the eccentricity should not be too high in order to prevent
passage in the Van Allen belts which could affect the on--board
capacitive reference sensor. Moreover, the orbital period $P_{\rm
OPT}$ should be shorter than the Earth's daily rotation of 24
hours. The orbital configuration proposed in \cite{lametal01}
would imply a semi major axis $a_{\rm OPT}=29300$ km and an
eccentricity $e_{\rm OPT}=0.478$. With such values the difference
of the gravitational potential $U$, which is relevant for the
gravitational redshift test, would amount to \eqi \frac{\Delta
U}{c^2}=\frac{GM_{\oplus}}{c^2
a}\left[\frac{1}{(1-e)}-\frac{1}{(1+e)}\right]\sim 1.8\times
10^{-10},\lb{reds} \eqf where $G$ is the Newtonian gravitational
constant, $M_{\oplus}$ the mass of Earth and $c$ the speed of
light in vacuum. The result of \rfr{reds} is about three orders of
magnitude better than that obtainable in an Earth--based
experiment.

An essential feature of OPTIS is the drag--free control of the
orbit. Drag--free motion is required for the special and general
relativistic tests which are carried through using optical
resonators. Even very small residual accelerations of $10^{-7}\,
g$ may distort the resonators leading to error signals. As a
by--product, this drag--free control also guarantees a very high
quality geodesic motion which may be used, when being tracked, as
probe of orbital relativistic gravitational effects.

For a drag--free motion of the satellite a sensor measuring the
actual acceleration and thrusters counteracting any acceleration
to the required precision are needed. The sensor, which is based
on a capacitive determination of the position of a test mass, has
a sensitivity of up to $10^{-12} \hbox{cm}\ {\hbox{\rm
s}}^{-2}/\sqrt{\hbox{Hz}}$ \cite{toub01}.
%This means that for one
%orbit of about 12 h the difference of the real position from the
%position achieved by ideal drag--free motion is of the order of 2
%mm.
%Similar drag--free systems of similar
%accuracy and with mission adapted modifications will be used, for
%example, in the MICROSCOPE \cite{micro} and STEP \cite{step}
%space--based missions aimed to testing the weak equivalence
%principle to a 10$^{-15}$--10$^{-18}$ level.
These systems have a
lifetime of many years.
%----------------------------------------------------------------
\subsection{The compatibility of OPTIS with the gravitomagnetic
tests} In this contribution we wish to show that the rather free
choice of the orbital parameters of OPTIS and the use of a new
drag--free technology open up the possibility to extend its
scientific significance with new important general relativistic
gravitomagnetic tests as well.
%Ciufpar
Indeed, it would be of great impact and scientific significance to
concentrate as many relativistic tests as possible in a single
mission, including also measurements in geodesy, geodynamics and
Earth monitoring. Another important point is that OPTIS is
currently under serious examination by a national space agency-the
German DLR. Then, even if it turns out that OPTIS would yield
little or no advantages for the measurement of the Lense--Thirring
effect with respect to the originally proposed LARES, if it will
be finally approved and launched it will nevertheless be a great
chance for detecting, among other things, the Lense-Thirring
effect. The main characteristics of such a mission are the already
mentioned drag--free technique for OPTIS and the SLR technique for
tracking. Today it is possible to track satellites to an accuracy
as low as a few mm. This may be further improved in the next
years.

It seems that an orbital configuration of OPTIS identical to that
of LARES of Table 1 would not be in dramatic contrast with the
requirements for the other originally planned tests. For example,
with the LARES orbital configuration the difference in the
gravitational potential $\frac{\Delta U}{c^2}$ would be of the
order of $3\times 10^{-11}$, which is only one order of magnitude
smaller than the one that could be obtained with the originally
proposed OPTIS configuration.

By assuming for OPTIS the same orbital configuration of LARES the
following combination yields high accuracy \eqi \dot\Omega^{\rm
LAGEOS}+c_1\dot\Omega^{\rm LAGEOS\ II}+c_2\dot\Omega^{\rm
OPTIS}+c_3\dot\omega^{\rm OPTIS} \ =61.8\mu_{\rm LT
},\lb{combinopg}\eqf with \eqi c_1  \sim 3\times 10^{-3},\ c_2
\sim  9.9\times 10^{-1},\ c_3  \sim  1\times 10^{-3}\lb{ccF3}.
\eqf The parameter $\mu_{\rm LT}$ is 0 in Galileo--Newton
mechanics and 1 in GTR.  The resulting relativistic signal would
be a linear trend with a slope of 61.8 mas yr$^{-1}$. The
combination of \rfr{combinopg} cancels out the first three even
zonal harmonics so that the systematic error due to the remaining
harmonics of higher degree amounts to $\left({\delta\mu_{\rm
LT}}/{\mu_{\rm LT}}\right)_{\rm even\ zonals}=3\times
10^{-3}\lb{erznopg}.$ The variance matrix of EGM96 up to degree
$l=20$ has been used. It can be shown that this result is
insensitive to the orbital injection errors in the inclination of
OPTIS.

With regard to the non-gravitational perturbations on the LAGEOS
satellites, only the contributions of the nodes of LAGEOS and
LAGEOS II, weighted by the small coefficients of \rfr{ccF3}, have
to be considered. This is quite relevant in the final error budget
because, as already pointed out, the nodes of the LAGEOS
satellites, contrary to the perigees of these laser-ranged
satellites, are orbital elements much less sensitive to the action
of the non--gravitational perturbations. With regard to the effect
on the non--gravitational perturbations on the OPTIS satellite, it
turns out that, by assuming a residual unbalanced acceleration of
$10^{-12}$ cm s$^{-2}$, the impact of the node of OPTIS would be
of the order of $10^{-3}$ and that of the perigee of OPTIS would
be of the order of $10^{-4}$ in view of the coefficients of
\rfr{ccF3}. So, according also to the evaluations of Table 2 and
Table 3 of \cite{iorlucciuf02}, over $T_{\rm obs}$ = 7 years
$\left({\delta\mu_{\rm LT}}/{\mu_{\rm LT}}\right)_{\rm NGP}\sim
3\times 10^{-3}.$
%It should be noted that
%the estimate of \rfr{ngnpg} is probably pessimistic. Indeed, the
%periods of many time--dependent perturbations of the nodes of
%LAGEOS and LAGEOS II, contrary to the perigee of LAGEOS II, are
%far shorter than 7 years\footnote{For example, the period of the
%tesseral tidal perturbation $K_1$, which is one of the most
%powerful perturbations affecting the node of LAGEOS, amounts to
%2.85 years \cite{iortid01} .}, so that would be possible to adopt
%a $T_{\rm obs}$ of just a few years\footnote{It would allow to
%save fuel needed for the drag--free apparatus.} during which it
%should be possible to fit and remove the small time--varying
%non-gravitational signals affecting the nodes of the LAGEOS
%satellites or average them out.
%to  $10^{-5}$
Last but not least, note that the impact of the perigee of LAGEOS
II, difficult to be modelled at a high level of accuracy, is
absent.
%the potentially unpredictable impact of
%the perigee of LAGEOS II is absent, contrary to \rfr{erzong}.

So, the total final systematic error budget in measuring the
Lense--Thirring effect with \rfr{combinopg} should be better than
$1\%$.
%-----------------------------------------------------------------
%\subsection{The problem of the eccentricity}

Perhaps the major point of conflict between the original designs
of the OPTIS and LARES missions is represented by the eccentricity
$e$ of the orbit of the spacecraft. Indeed, while for the
gravitational redshift test, given by \rfr{reds}, a relatively
large value of $e$  is highly desirable, the originally proposed
LARES orbit has a smaller eccentricity. The point is twofold: on
one hand, it is easier and cheaper, in terms of requirements on
the performances of the rocket launcher, to insert a satellite in
a nearly circular orbit, and, on the other, the present status of
the ground segment of SLR would assure a uniform tracking of good
quality for such kind of orbits. In fact, the large eccentricity
of the originally proposed OPTIS configuration, contrary to the
other existing geodetic satellites of LAGEOS--type, would not
allow for a uniform coverage of the laser--ranged data in the
sense that certain portions of the orbit might remain poorly
tracked.

However, the originally proposed OPTIS mission implies the use of
a ARIANE 5 rocket to insert the spacecraft into a GTO orbit and,
then, the use of a kick motor. Moreover, it may be reasonable to
assume that, when OPTIS/LARES will be launched, the network of SLR
ground stations will have reached a status which will allow to
overcome the problem of reconstructing rather eccentric orbits to
a good level of accuracy.

Then, a reasonable compromise between the OPTIS and LARES
requirements could be an eccentricity of, say, $e=0.1$. In that
case \rfr{reds} yields a gravitational redshift of $\frac{\Delta U
}{c^2}=7.3\times 10^{-11}$. With regard to the Lense--Thirring
effect, it turns out that the error due to the even zonal
harmonics of geopotential would amount, from \rfr{combinopg}, to
1.5$\%$, according to the variance matrix of EGM96 up to degree
$l=20$. Also in this case it would be insensitive to the orbital
injection errors in the inclination. However, the forthcoming
Earth gravitational models from CHAMP \cite{pav00} and GRACE
\cite{riesetal03} should greatly improve such estimates. With a
larger eccentricity the impact of the non--gravitational
perturbations would be reduced and, on the other hand, the
accuracy of the measurement of the various post--Newtonian
gravitational effects on the OPTIS/LARES perigee would be
increased.
%-------------------------------------------------------
\section*{Acknowledgments}
I am grateful to Prof. Klapdor--Kleingrothaus for his kind
invitation to the BEYOND2003 conference and to L. Guerriero for
his support while at Bari. Thanks also to the LARES--OPTIS team
members I. Ciufolini, E. Pavlis, D. Lucchesi, C. L${\rm \ddot
a}$mmerzahl, H. Dittus, S. Schiller.
%-----------------------------------------------------------

\end{document}